\documentclass{article}

\pdfoutput=1
\usepackage{arxiv}

\usepackage[utf8]{inputenc} 
\usepackage[T1]{fontenc}    
\usepackage{hyperref}       
\usepackage{url}            
\usepackage{booktabs}       
\usepackage{amsfonts}       
\usepackage{amsmath}

\usepackage{nicefrac}       
\usepackage{microtype}      
\usepackage{lipsum}
\usepackage{graphicx}
\usepackage[natbibapa]{apacite}
\graphicspath{ {./images/} }

\usepackage{graphicx}
\usepackage[ruled,vlined,noend]{algorithm2e}
\usepackage{subcaption}
\usepackage{float}
\usepackage{mwe}

\newcommand{\KL}{D_\mathrm{KL}}

\title{Relaxing the Constraints on Predictive Coding Models}

\author{
 Beren Millidge \\
  School of Informatics\\
  University of Edinburgh\\
  \texttt{beren@millidge.name}
   \And
 Alexander Tschantz \\
  Sackler Center for Consciousness Science\\
  School of Engineering and Informatics\\
  University   Sussex \\
  \texttt{tschantz.alec@gmail.com} \\
  \And
  Anil K Seth \\
  Sackler Center for Consciousness Science\\
   Evolutionary and Adaptive Systems Research Group\\
  School of Engineering and Informatics\\
  University of Sussex\\
  \texttt{A.K.Seth@sussex.ac.uk} \\
  \And
 Christopher L Buckley \\
  Evolutionary and Adaptive Systems Research Group\\
  School of Engineering and Informatics\\
  University of Sussex\\
  \texttt{C.L.Buckley@sussex.ac.uk} 
}

\begin{document}
\maketitle
\begin{abstract}
    Predictive coding is an influential theory of cortical function which posits that the principal computation the brain performs, which underlies both perception and learning, is the minimization of prediction errors. While motivated by high-level notions of variational inference, detailed neurophysiological models of cortical microcircuits which can implements its computations have been developed. Moreover, under certain conditions, predictive coding has been shown to approximate the backpropagation of error algorithm, and thus provides a relatively biologically plausible credit-assignment mechanism for training deep networks. However, standard implementations of the algorithm still involve potentially neurally implausible features such as identical forward and backward weights, backward nonlinear derivatives, and 1-1 error unit connectivity. In this paper, we show that these features are not integral to the algorithm and can be removed either directly or through learning additional sets of parameters with Hebbian update rules without noticeable harm to learning performance. Our work thus relaxes current constraints on potential microcircuit designs and hopefully opens up new regions of the design-space for neuromorphic implementations of predictive coding.
\end{abstract}

\section{Introduction}

Despite fulfilling a bewildering variety of functions, there is some evidence that the neocortex is surprisingly uniform \citep{rockel1980basic,carlo2013structural,hassabis2017neuroscience,hawkins2007intelligence,george2009towards}, although this claim is still controversial \citep{rakic2008confusing,herculano2008basic}. This potential uniformity has fuelled speculations that there may be a single general algorithm or canonical computation that the cortex performs. Such an algorithm would then be implemented in neural circuitry by stereotyped cortical microcircuits which are simply repeated many times throughout the cortex, and which perform the same core function. The observed functional specialisation of the cortex would be achieved primarily through differing inputs and initial developmental connectivity patterns rather than fundamental algorithmic differences between regions. One such potentially unifying theory of cortical function is predictive coding which is subsumed by the more general free-energy principle \citep{rao1999predictive,friston2003learning,friston2019free,friston2006free,friston2010free}. Predictive Coding combines a strong and elegant theoretical framework anchored in deep thermodynamic considerations \citep{friston2012free,friston2019free}, to a straightforward interpretation in terms of variational inference \citep{friston2003learning,buckley2017free}, while also being supported in its concrete predictions by a wide array of neurophysiological data \citep{walsh2020evaluating,keller2018predictive,bastos2012canonical,spratling2008reconciling,spratling2010predictive}. The basic idea of predictive coding is that instead of propagating all information through the network, it suffices to send the information that is not predicted -- prediction errors \citep{clark2013whatever}. Thus instead of the typical idea of a feedforward sweep conveying information about stimuli upwards, instead upward information is constantly meeting top-down predictions, and only the differences between these predictions and real sensory data -- the prediction error -- is transmitted upwards \citep{clark2015surfing,friston2003learning,rao1999predictive}.

Due to the recent successes of deep neural networks, the backpropagation of error algorithm (backprop) \citep{rumelhart1985feature,werbos1982applications} which is the central algorithm used to train deep neural networks, is also a strong contender for the place of unified brain theory \citep{marblestone2016toward,richards2019deep,hassabis2017neuroscience}. The core reason for the success of backpropagation is that it correctly assigns credit to (i.e. it computes the derivative of) each parameter in the network for its contribution to the final output of the network. This algorithm has been generally considered to be biologically implausible due to seemingly requiring the propagation of a large amount of non-local information through the network \citep{crick1989recent}. However, a small number of recent architectures have been proposed which can implement backpropagation or approximations thereunto with simple biologically plausible local learning rules \citep{whittington2017approximation,whittington2019theories,scellier2018generalization,bengio2017stdp,sacramento2018dendritic,ororbia2019biologically,millidge2020activation}. Interestingly, it has been shown that predictive coding networks, although proposed independently of backprop, can be seen as performing the backpropagation of error algorithm \citep{whittington2017approximation}, even on arbitrary computation graphs \citep{millidge2020predictive}, and it does so in a very natural way which circumvents the nonlocality of error propagation by postulating intermediate error neurons. These error neurons, although they use only local information, encode enough information to allow the correct derivatives to be backpropagated through the network over the course of multiple dynamical iterations. Moreover, predictive coding allows networks to be trained which are both generative and discriminative classifiers, which can learn in an unsupervised or a supervised fashion, and which can be trained and tested without the need for multiple backwards phases. Predictive coding thus provides a robust and flexible biologically plausible scheme for approximating backprop, possesses a number of neural process theories showing how it may be implemented in cortical microcircuitry, and also has a straightforward interpretation in terms of variational inference, thus tying it nicely into theories of the Bayesian Brain \citep{knill2004bayesian,pouget2013probabilistic,doya2007bayesian}. As a theory, predictive coding is perhaps unique in how it straddles all three levels of Marr's hierarchy. Computationally, it proposes that the brain implements a variational approximation to Bayesian inference, and does so under Gaussian and Laplacian assumptions over the generative model and variational density respectively \citep{friston2005theory,buckley2017free,millidge2019combining,millidge2019implementing}. Algorithmically, predictive coding theory provides biologically plausible Hebbian update rules which can be straightforwardly implemented in a hierarchical network of layers, and in terms of biological hardware, predictive coding makes many detailed predictions about the exact form of the cortical microcircuits required, and several fully-fledged process theories exist \citep{bastos2012canonical,shipp2016neural,shipp2013reflections,kanai2015cerebral,keller2018predictive,walsh2020evaluating,walsh2020evaluating}.

While predictive coding possesses several theorised neural implementations, and in this paper we remain largely agnostic between them, there are still several general aspects of the scheme that are heavily restrictive if not completely biologically implausible. In this paper, we take aim at these remaining biological implausibilities and try to dissolve them if possible. It is worth noting here that biological plausibility is an amorphous term. In the literature, it is often taken to be some combination of Hebbian update rules, using only information (theoretically) available locally at the synapse, and requiring patterns of connectivity which do not need to be tightly coordinated and are robust to dropped or sparse connections. Biological plausibility is nevertheless a fundamentally abstract notion, in that all biologically plausible process theories that try to approximate backprop generally shy away from the precise details of neuroanatomy and synaptic function. The 'neurons' in these models are often simply rate-coded integrate and fire neurons with scalar synaptic connection weights, and thus these theories exist at a level of abstraction above neural spike-trains and the biological details of synaptic plasticity. We continue in that theoretical tradition here.

We focus on three outstanding issues of biological plausibility in predictive coding networks. The first is the problem of weight symmetry, or the requierd equality of forward and backward weights, which also affects backprop. The learning rules in these networks require information to be sent 'backwards' through the network. Since synaptic connections are generally assumed to be uni-directional, in practice this means that these backward messages need to be sent through a second set of backwards connections with the exact same synaptic weights as the forward connections. Clearly, expecting the brain to have an identical copy of forward and backward weights is infeasible. In the literature this is often called the weight-transport problem \citep{crick1989recent}. In this paper, we show that this problem can be solved in predictive coding networks through randomly initialized backwards weights trained with a separate Hebbian learning rule, which also only requires local information. This removes the necessity of beginning with symmetrical or identical weights and proposes a biologically plausible method of learning good backward weights from scratch in an unsupervised fashion.

There is also a small literature on addressing the weight transport problem in deep neural networks trained with backprop. A key paper \citep{lillicrap2014random,lillicrap2016random} shows that simply using random backwards weights is sufficient for some degree of learning. This method is called feedback alignment (FA) since during training the feedforward weights learn to align themselves with the random feedback weights so as to transmit useful gradient information. A variant of this -- direct feedback alignment (DFA) \citep{nokland2016direct} has been shown that direct forward-backward connectivity is not necessary for successful learning performance. Instead, all layers can receive backwards feedback directly from the output layer. It has also been shown \citep{liao2016important} that performance with random weights is substantially improved if the forward and backward connections share merely the same sign, which is less of a constraint than the exact value. One further possibility is to learn the backwards weights with an independent learning rule. This has been proposed independently in \citep{amit2019deep} and \citep{akrout2019deep} who initialize the backwards weights randomly, but train them with some learning rule. Our work here differs primarily in that we show that this learning rule works for predictive coding networks while they only apply it to deep neural networks learnt with backprop. Moreover, our Hebbian learning rule can be straightforwardly derived in a mathematically principled manner as part of the overarching variational framework of predictive coding. 

The second problem is one of backward nonlinear derivatives. In predictive coding networks (along with backprop), the update and learning rules require the pointwise derivatives of the activation function to be computed at each neuron. For individual biological neurons, while a nonlinear forward activation function is generally assumed, the ability to compute the derivative of the activation function is not known to be straightforward. While in some cases this issue can be ameliorated by a judicious choice of activation function -- for instance the pointwise derivative of a a rectified linear unit is simply 0 or 1, and is a simple step function of the firing rate -- the problem persists in the general case. In this paper, we show that, somewhat surprisingly, it is possible to simply ignore these pointwise derivatives with relatively little impact on learning performance, despite the update rules now being mathematically incorrect. This may free the brain of the burden of having to compute these quantities.

A third issue, specific to frameworks that explicitly represent prediction errors, is the requirement of one-to-one connections between activation units and their corresponding error units. While not impossible, this precise, one-to-one connectivity pattern is likely difficult for the brain to create and maintain throughout development and learning. In this paper, we show that learning can continue unaffected despite random connectivity patterns between value and error units as long as these connection weights can also be learned. We propose a further Hebbian and biologically plausible learning rule to update these weights which also only requires local information. Finally, we experiment with combining out solutions to all of these problems together to produce a fully relaxed predictive coding architecture. Importantly, this architecture possesses a simple bipartite but otherwise fully connected connectivity pattern with separate learnable weight matrices covering every connection, all of which are updated with local Hebbian learning rules. We show that despite the simplicity of the resulting relaxed scheme, that it can still be trained to high classification accuracy comparable with standard predictive coding networks and ANNs using backpropagation.

In sum, we directly address and solve three outstanding issues of biological plausibility in predictive coding networks. We thus demonstrate that the predictive coding learning algorithm is robust to major changes such as imperfect backwards weights, as well as to some degree of mathematical incorrectness due to the removal of the nonlinear derivatives. Our study provides further evidence that predictive coding is suitably robust and distributed to be able to be implemented in the brain, and we remove several significant constraints on building neuroarchitectural models of predictive coding in the cortex. Finally, it is worth stressing, that although we only address these issues in the context of predictive coding networks, many other frameworks for biologically plausible learning, such as target-prop, and equilibrium-prop also impose similar implausible conditions, and it would be an interesting avenue for future work to see if our remedies also apply to those algorithms.

\section{Predictive Coding Networks}

Predictive coding has a long history in signal processing, where it subsumes delta-encoding, a form of compression widely used in engineering and computer science \citep{makhoul1975linear,vaseghi2008advanced}. It was first proposed to explain computations in the retina \citep{srinivasan1982predictive}, then was applied to understand hierarchical processing in the visual cortex \citep{rao1999predictive}, and has since been extended into a general theory of brain function \citep{friston2003learning,friston2005theory,friston2008hierarchical}, with deep connections to other theories of brain function such as biased competition theory \citep{spratling2008reconciling}. Predictive coding possesses several neurobiologically realistic process theories \citep{bastos2012canonical,kanai2015cerebral,shipp2013reflections,shipp2016neural} and, under certain conditions, is known to approximate the backpropagation of error algorithm \citep{whittington2017approximation,millidge2020predictive}.

While predictive coding can be motivated directly from neurophysiological considerations, it also has an elegant theoretical interpretation as a variational inference algorithm. This situates predictive coding firmly within the Bayesian Brain framework \citep{knill2004bayesian}, by positing that the core algorithm the cortex is doing is approximate (variational) Bayesian inference. Variational methods proceed by approximating the true posterior density by a variational density that is under the control of the modeller. This is done by deriving a tractable quantity called the variational free energy, which is an upper bound on the log model evidence, and equal up to a constant to the divergence between true and approximate posteriors. By minimizing the variational free-energy, denoted $\mathcal{F}$, this divergence is minimized and the approximate posterior becomes a successively better approximation to the true posterior.

Under Gaussian assumptions about the generative model, and Laplacian assumptions on the variational density -- see \citep{buckley2017free,bogacz2017tutorial,millidge2019combining,millidge2019implementing} -- the predictive coding update rules can be derived straightforwardly as a gradient descent on $\mathcal{F}$ relative to the variational parameters $\mu$ and the parameters of the generative model $W$ (see Appendix A for a full derivation of the free-energy functional and the update rules for all the parameters). This perspective is very useful since it allows for arbitrarily extending either the variational inference or generative model with additional parameter sets, and then deriving the update rules for these parameters is as simple as computing gradients with respect to $\mathcal{F}$. Moreover, this perspective provides great insight into what predictive coding is \emph{doing} in an algorithmic sense -- it is trying to maximize the log model evidence of its observations by inferring the states and parameters of a hierarchical generative model of the world.

\begin{figure}%
    \centering
    \subfloat[\centering `Standard' Generative PC]{{\includegraphics[width=0.40\textwidth]{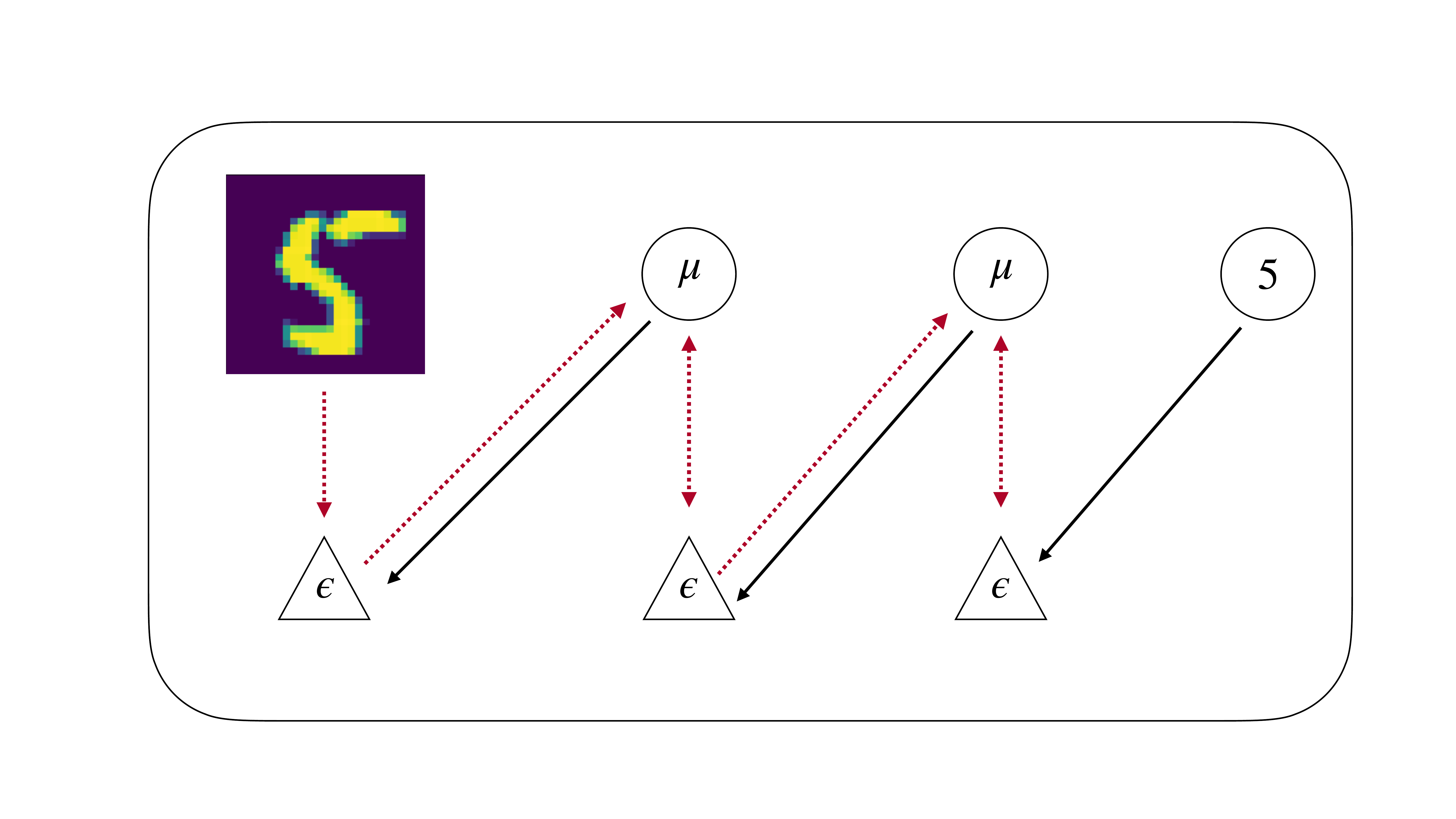} }}%
    \qquad
    \subfloat[\centering `Reverse' Discriminative PC ]{{\includegraphics[width=0.40\textwidth]{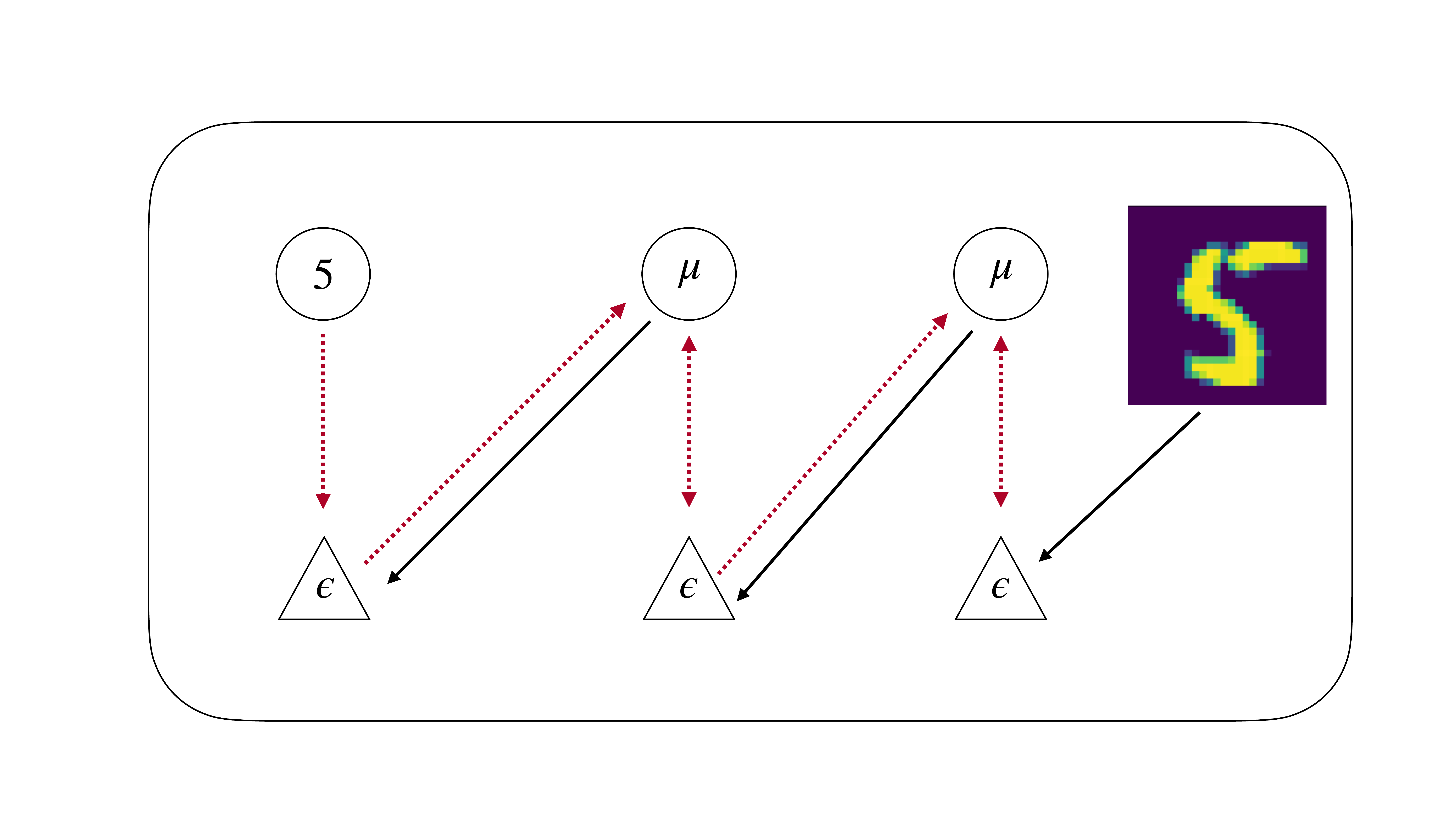} }}%
    \caption{Schematic architectures for the a.) Standard, or generative predictive coding setup, or b.) Reverse, or discriminative architecture. There are two types of neurons in a predictive coding network -- 'value neurons' or $\mu$s which maintain representations of expected stimuli, and 'error nodes' $\epsilon$ which compute the prediction errors. In the generative model, the image input (in this case an MNIST digit) is presented to the bottom layer of the network, and the top layer is fixed to the label value (5). Predictions (in black) are passed down and prediction errors (in red) are passed upwards until the network equilibrates. In the discriminative mode, the input image is presented to the top of the network and the label is presented at the bottom. Thus the network aims to `generate' the label from the image. The top-down flow of predictions becomes analogous to the forward pass in an artificial neural networks, and the bottom-up prediction errors become equivalent to the backpropagated gradients.}%
\end{figure}

Predictive coding gives rise to a straightforward neurobiological scheme which can model hierarchical layers in the cortex. Neural models of predictive coding typically maintain two separate populations of neurons -- value neurons, which encode estimates of sensory data, and error neurons which encode prediction errors. Predictions are generated from the value neurons of the layer above projected downward through a weight matrix. Prediction errors are simply the difference between the activations of the value units at a specific level and the top down predictions. Neurally, this is typically implemented through inhibitory top-down connections and excitatory driving connections from value neurons to error neurons \citep{bastos2012canonical}, although other implementations are possible \citep{spratling2008reconciling}.

This scheme can be formally described according to the following rules \footnote{It is important to note that predictive coding networks as defined in \cite{bogacz2017tutorial,friston2003learning} also typically have \textit{precision} parameters $\Pi$ which are inverse variances which can weight the prediction errors by their reliability. In general throughout we ignore precisions in what follows -- we assume $\Pi = \mathrm{I}$. \citep{bogacz2017tutorial} propose a biologically plausible implementation of precision using an additional set of inhibitory neurons.}.
\begin{align}
    \epsilon^l &= \mu^l - f(\theta^{l+1} \mu^{l+1}) \\
    \frac{d\mu^l}{dt} &= -\epsilon^l + {\theta^l}^T \epsilon^{l-1} f'(\theta^l \mu^l) \\
    \frac{d\theta^l}{dt} &= \epsilon^{l-1} f'(\theta^l \mu^l) {\mu^l}^T
\end{align}

Equation 1 states that prediction errors are computed as a simple subtraction of the value neurons at a layer and the prediction from the layer above. The vector $\mu^l$ represents the activity of the value neurons at a specific level $l$. The vector $\epsilon^l$ is a vector of the activity of the error neurons at a level l. Predictions are mapped down from the higher layers through a set of weights, denoted $W$ which is an MxN matrix where M is the number of neurons at level $l$ and N is the number of neurons at level $l+1$. $f(x)$ is a nonlinear activation function applied to the outputs of a neuron and $f'(x) = \frac{\partial f(x)}{\partial x}$ is the pointwise derivative of the activation function. 

Equation 2 specifies the update rule for the $\mu^l$ at a specific layer. The update is equal to the sum of the prediction errors projected up from the layer below, multiplied by the top down predictions and projected back through the weight matrix and subtracted from the prediction errors at the current layer. This is a biologically plausible learning rule as it is a simple sum of multiplication of locally available information. Note: the the update includes prediction error terms from both the current layer and the layer below. This equation is why it is necessary to transmit prediction errors upwards.

Equation 3 is the update rule for the weights $W$. This obeys Hebbian plasticity since it is simply a multiplication of the two quantities available at each end of the synaptic connection -- the prediction error of the layer below and the value neurons at the current layer. The only slight difficulty is the derivative of the nonlinear activation function of the prediction. While this information is locally available in principle, it requires a somewhat more complex neural architecture and it is not certain that the derivatives of activation functions can be computed straightforwardly by neurons. Luckily, we show below that this term is not needed for successful operation of the learning rule.

Predictive coding networks can operate in both supervised and unsupervised modes. Training in the supervised case (as considered in this paper), occurs by fixing the top level of the network to the label and the bottom layer to the input. Predictions and prediction errors are propagated through the network and it is allowed to settle into equilibrium. Then the weights are updated. Typically, the $\mu$s are updated first (Equation 2) to convergence and then the weights are updated once (Equation 3). This separation of timescales represents the distinction between (fast) inference and (slow) learning. In the supervised mode the input is presented to the bottom layer of the network, but the value neurons at the top are allowed to update freely to convergence. These values can then be read out as a prediction of the label.

\section{Results}
To test the performance of the predictive coding network under various relaxations, we utilize the canonical MNIST and FashionMNIST benchmark datasets from machine learning. Since this is a supervised classification task, we follow the approach of \cite{whittington2017approximation} and \cite{millidge2020predictive}, who utilized a `reverse' predictive coding architecture where the inputs were presented to the top layer of the network and the labels were predicted at the bottom. This formulation allows for the straightforward representation of supervised learning problems in predictive coding. In effect the networks tries to `generate' the label from the image. We utilized a 4-layer predictive coding network consisting of 784,300,100, and 10 neurons in each layer respectively. We tested both rectified-linear (relu) and hyperbolic tangent (tanh) activation functions.  During training the $\mu$s were updated for 100 iterations using Equation 2 with both the input and labels held fixed. After the iterations of the $\mu$s, the remaining prediction errors in the network were used to update the weights according to Equation 3. 
At test time, a digit image was presented to the network, and the top-down predictions of the network were propagated downwards to produce a prediction at the lowest layer, which was compared to the true label to obtain the test accuracy. 

The network was trained and tested on the MNIST and FashionMNIST datasets. MNIST is a dataset of 70,000 28x28 grayscale images of handwritten digits from 0 to 9. The goal is to predict the digit from the image. The FashionMNIST dataset contains images of different types of clothing, which must be sorted into ten classes. FashionMNIST is a more challenging classification dataset, while also serving as a drop-in replacement for MNIST since its data is in exactly the same format. The input images were flattened into a 784x1 vector before being fed into the network. Labels were represented as one-hot vectors and were smoothed using a value of 0.1 for the incorrect labels. The dataset was split into a training set of 60000 and a test set of 10000 images. All weight matrices were initialized as draws from a multivariate gaussian with a mean of 0 and a variance of 0.05. It is likely, given the large literature on how to best initialize deep neural networks, that there exist much better initialization schemes for predictive coding networks as well, however we did not investigate this in this paper. All results presented in this paper were averaged over 5 random seeds. We plot error bars around the means as the standard deviation of the seeds.

\subsection{Weight Transport}

\begin{figure}[h]
    \label{Backward_Weight_Diagram}
   \begin{center}
   \includegraphics[width=13cm, height=8cm]{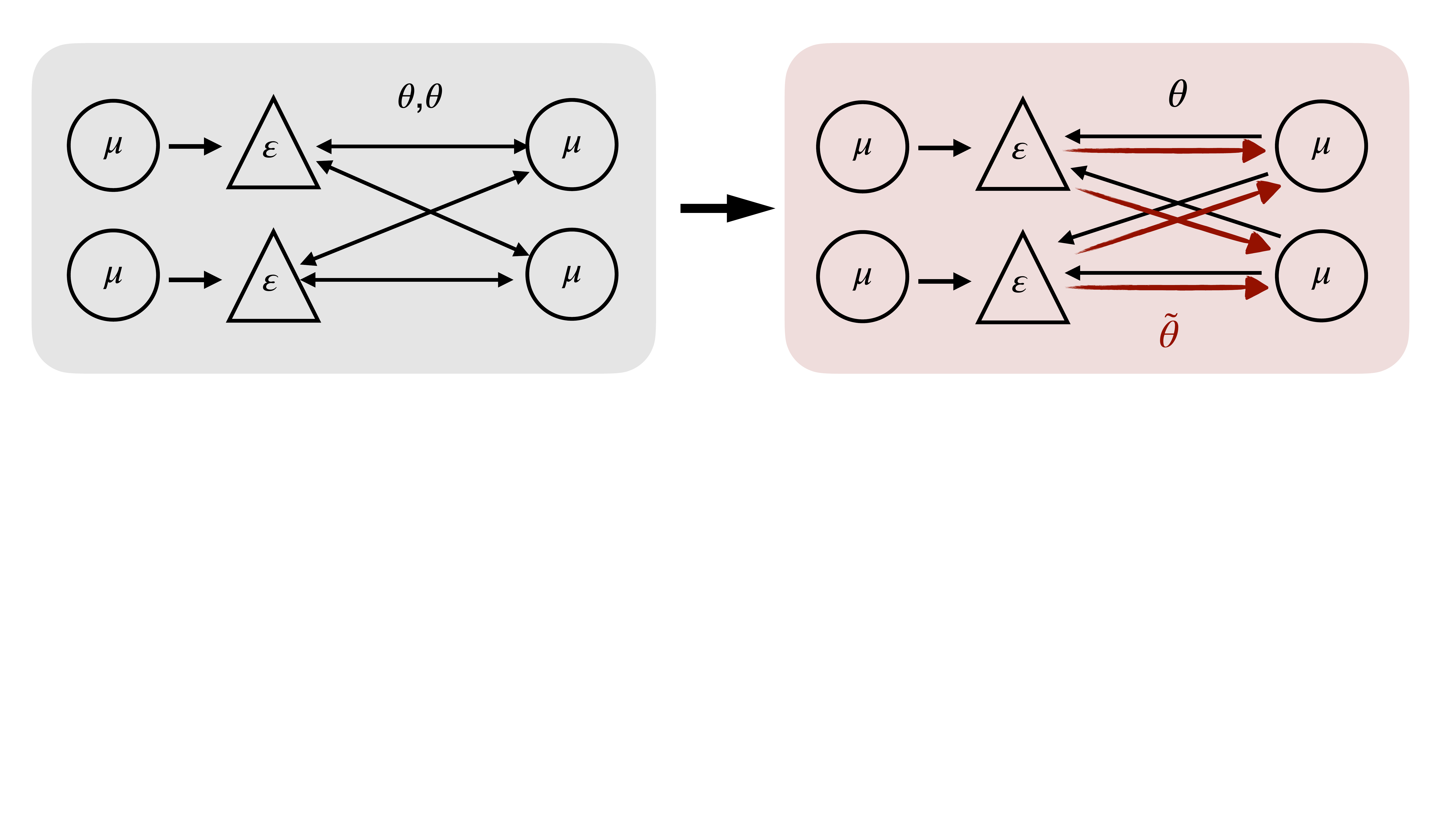}
   \end{center}
   \vspace{-3.9cm}
   \caption{The weight transport problem and our solution. On the left is the standard predictive coding network architecture. Our diagram represents the prediction errors $\epsilon$ of one layer receiving predictions and transmitted prediction errors to the value neurons of the layer above. Prediction errors are transmitted upwards using the same weight matrix $\theta^T$ as the predictions are transmitted downwards. On the right, our solution eschews this biological implausibility by proposing a separate set of backwards weight $\tilde{W}$ (in red), which are learned separately using an additional Hebbian learning rule.}
   \end{figure}

Mathematically, the weight transport problem is caused by the $\theta^T$ term in Equation 2. In neural circuitry this weight transpose corresponds to transmitting the message backwards through the same connections or, alternatively, an identical copy of the backward weights. We wish to replace this copy of the forward weights with an alternative, unrelated set of weights $\tilde{W}$. Unlike FA or DFA methods, which simply use random backwards weights, we propose to learn the backwards weights through a simple synaptic plasticity rule. 
\begin{align}
    \frac{d\tilde{\theta^l}}{dt} = \mu^l (f'(\theta^l \mu^l) \epsilon^{l-1})^T
\end{align}
This rule is Hebbian since it is just the multiplication of the activities of the units at each end of the connection. The backwards pointwise derivative poses a slight problem in that it is first multiplied with the errors of the level below. However, as we show below, pointwise nonlinear derivatives are not actually needed for good learning performance, so the problem is surmounted. We derive this rule from the free-energy in Appendix A.

This procedure allows us to begin with a randomly initialized set of backwards weights, but by applying the learning rule to these weights allows us to very quickly recover performance equal to the identical backwards weights. As shown in Figure 3, performance both with and without the learnt backwards weights is almost identical for both the relu and tanh nonlinearities and the MNIST and FashionMNIST datasets, thus suggesting that this approach of simply learning an independent set of backwards weights is a highly effective and robust method for tackling the weight transport problem.

\begin{figure}[ht] 
  \begin{subfigure}[b]{0.5\linewidth}
    \centering
    \includegraphics[width=0.75\linewidth]{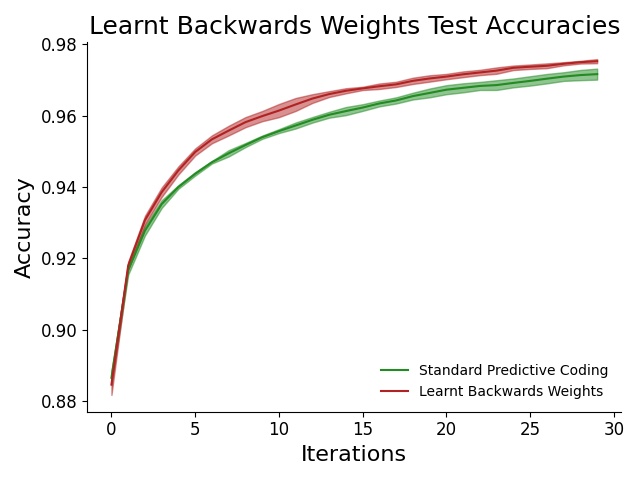} 
    \caption{\small MNIST dataset; tanh activation} 
    \vspace{4ex}
  \end{subfigure}
  \begin{subfigure}[b]{0.5\linewidth}
    \centering
    \includegraphics[width=0.75\linewidth]{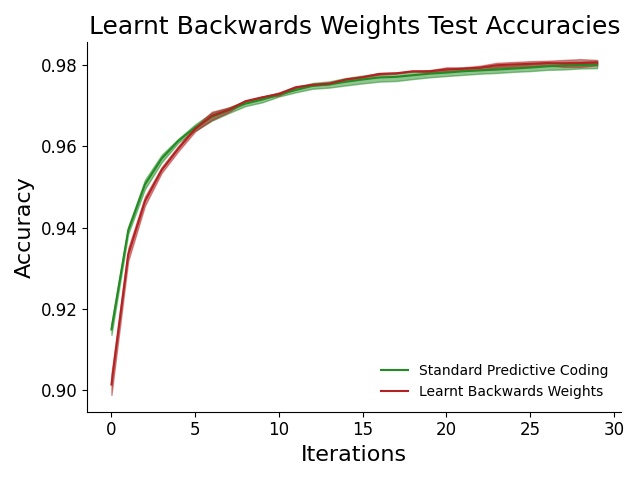} 
    \caption{\small MNIST dataset; relu} 
    \vspace{4ex}
  \end{subfigure} 
  \begin{subfigure}[b]{0.5\linewidth}
    \centering
    \includegraphics[width=0.75\linewidth]{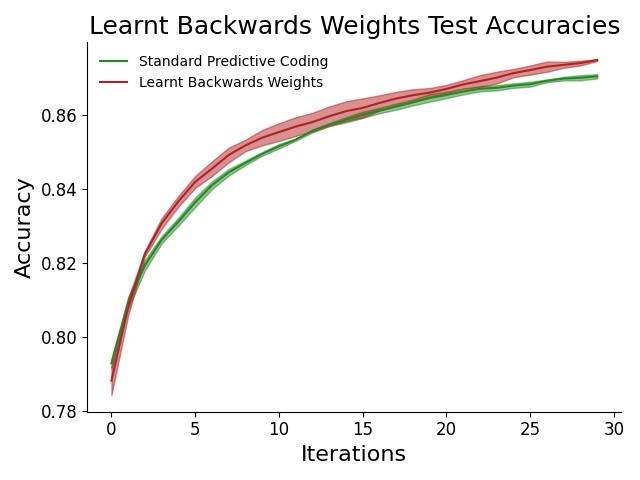} 
    \caption{\small Fashion dataset; tanh} 
  \end{subfigure}
  \begin{subfigure}[b]{0.5\linewidth}
    \centering
    \includegraphics[width=0.75\linewidth]{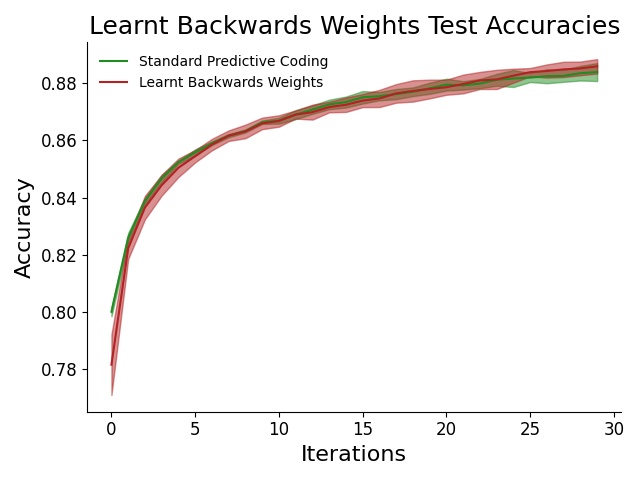} 
    \caption{\small Fashion dataset; relu} 
  \end{subfigure} 
  \caption{Test accuracy of predictive coding networks with both learnt backwards weights, and the ideal weight transposes with both relu and tanh activation functions on the MNIST and FashionMNIST datasets. Both networks obtain almost identical learning curves, thus suggesting that learnt backwards weights allow for a solution to the weight-transport problem.}
\end{figure}

 \subsection{Backwards nonlinear derivatives}

The second remaining biological implausibility is that of the backwards nonlinear derivatives. Note that in Equations 2 and 3, an f' term regularly appears denoting the pointwise derivative of the nonlinear activation function. Since these are pointwise derivatives, when the mathematics is translated to neural circuitry, these derivatives need to be computed at each individual neuron. It is not clear whether neurons are capable of easily computing with the derivative of their own activation function. We apply a straightforward remedy to address this. We simply experiment with removing the pointwise derivatives from the update rules. For instance Equation 2 would become just $\frac{d\mu^l}{dt} = -\epsilon^l + {\theta^l}^T \epsilon^{l-1}$. Perhaps surprisingly we found that this modification, although not mathematically correct, did little to impair performance of the model at classification tasks, except perhaps for the hyperbolic tangent nonlinearity in the FashionMNIST case.

\begin{figure}[ht] 
  \begin{subfigure}[b]{0.5\linewidth}
    \centering
    \includegraphics[width=0.75\linewidth]{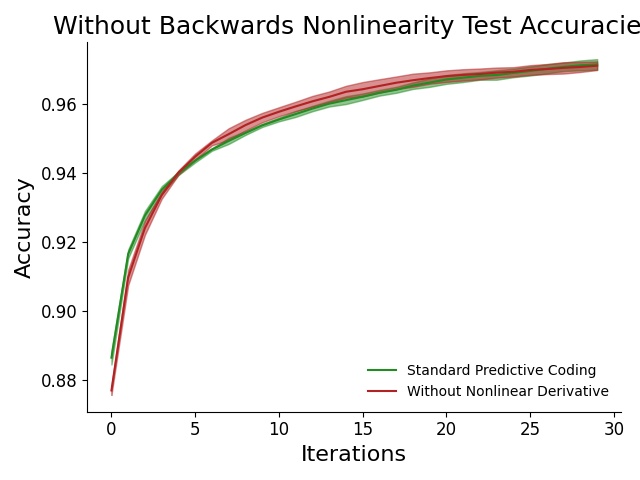} 
    \caption{\small MNIST dataset; tanh activation} 
    \vspace{4ex}
  \end{subfigure}
  \begin{subfigure}[b]{0.5\linewidth}
    \centering
    \includegraphics[width=0.75\linewidth]{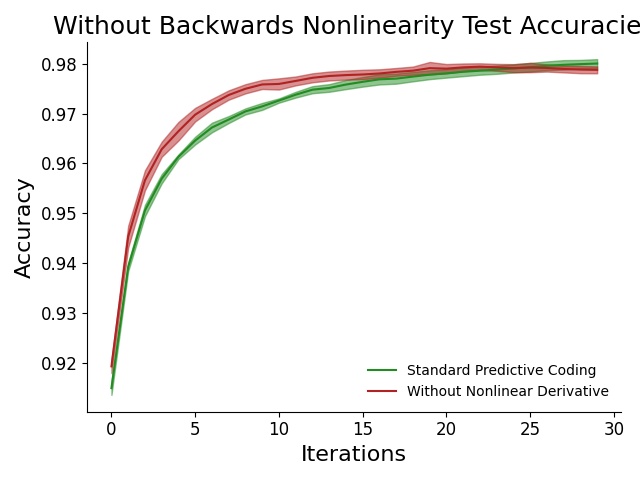} 
    \caption{\small MNIST dataset; relu activation} 
    \vspace{4ex}
  \end{subfigure} 
  \begin{subfigure}[b]{0.5\linewidth}
    \centering
    \includegraphics[width=0.75\linewidth]{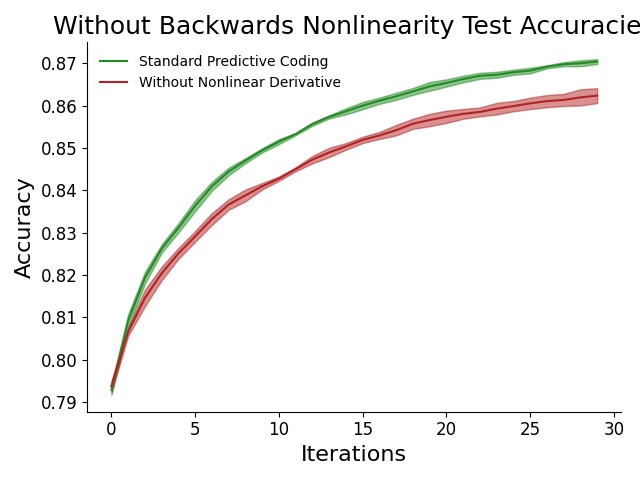} 
    \caption{\small Fashion dataset; tanh activation} 
  \end{subfigure}
  \begin{subfigure}[b]{0.5\linewidth}
    \centering
    \includegraphics[width=0.75\linewidth]{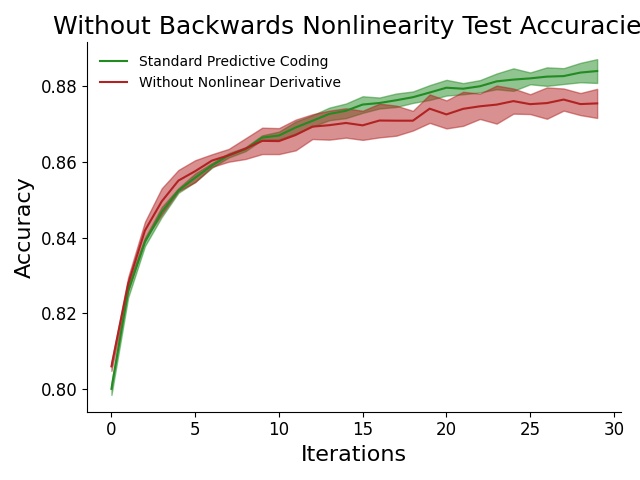} 
    \caption{\small Fashion dataset; relu activation} 
  \end{subfigure} 
  \caption{Test accuracy of predictive coding networks with and without the nonlinear derivative term, using relu and tanh activation functions on the MNIST and FashionMNIST datasets. We find that on the MNIST dataset performance is similar, while on the FashionMNIST dataset and the tanh activation function, the lack of the nonlinear derivative appears to slightly hurt performance.}
\end{figure} 
   
In effect, by removing the pointwise nonlinear derivatives, we have made the gradient updates linear in the parameters. Since the real updates are nonlinear, our update rules are simply the projection of the nonlinear update rules onto a linear subspace. However, using a similar argument to that in feedback alignment, we hypothesize that it is likely that the linear projection of the nonlinear updates are quite close in angle to the nonlinear updates, so the direction of the linear gradient, averaged over many batches and update steps, is sufficiently close to the true gradient as to allow for learning in this model.

\subsection{Error connections}
 
\begin{figure}
    \label{Error_Connection_Diagram}
   \begin{center}
   \includegraphics[width=16cm, height=8cm]{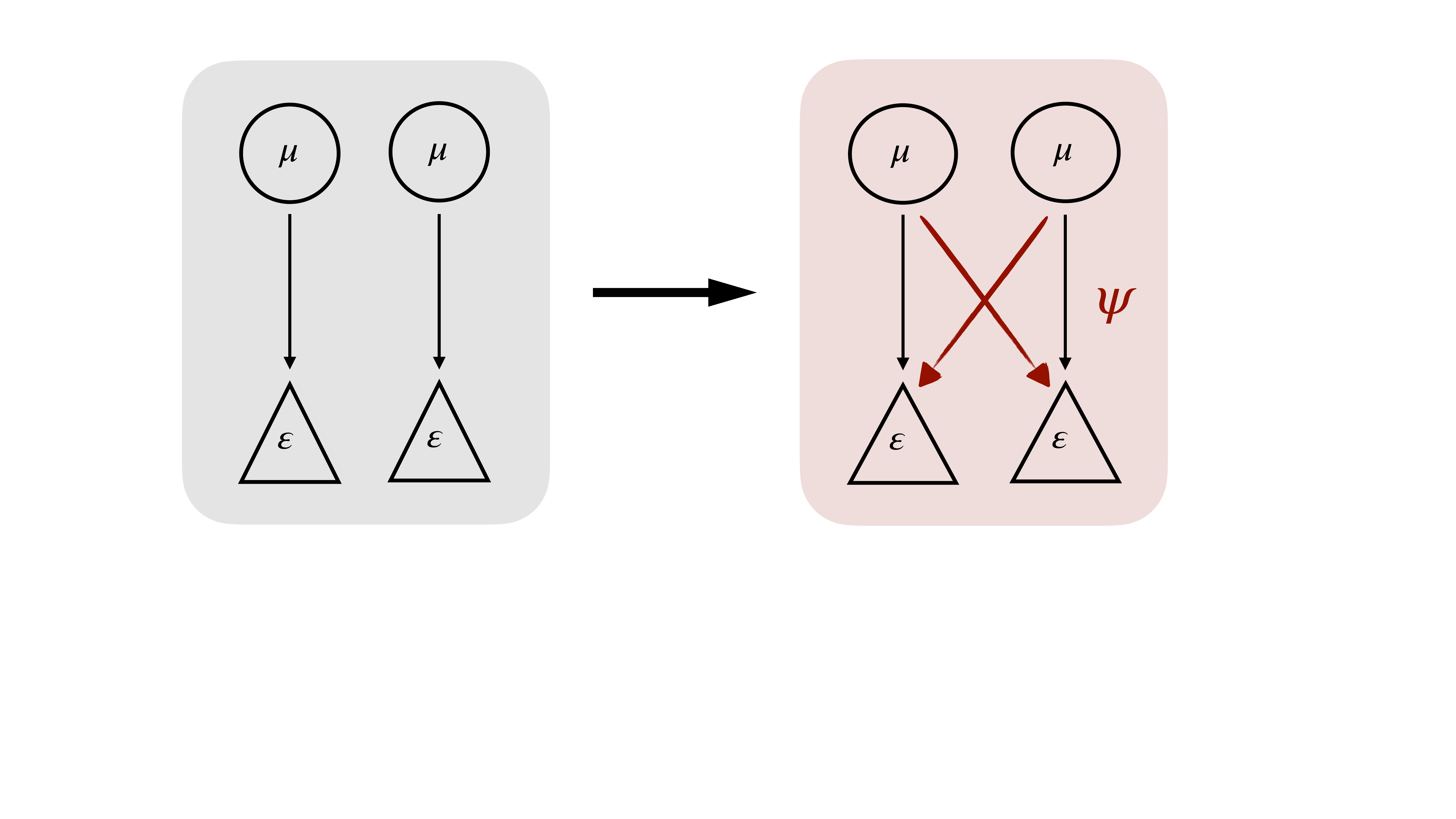}
   \end{center}
   \vspace{-3cm}
   \caption{The error-connectivity problem and our solution. On the left, the biologically implausible one-to-one connectivity between value and error nodes required by the standard predictive coding theory. On the right, our solution to replace these one to one connections by a fully connected connectivity matrix $\psi$. By learning $\psi$ with a Hebbian learning rule we are able to achieve comparable performance to the one-to-one connections with a fully dispersed connectivity matrix.}
   \end{figure}

The third and final biological implausibility that we address in this paper is that of the one-to-one connections between value and the error units at a given layer. This can be seen directly from Equation 1, but broken down into individual components (or neurons).
\begin{align}
    \epsilon_i^l = \mu_i^l - f(\sum_j \theta^{l+1}_{i,j} \mu^{l+1}_{j}) 
\end{align}

We see that the activity of the error unit vector (i.e. each error neuron $\epsilon_i$) is driven by a one-to-one connection from its matching value neuron $\mu_i$. By contrast, the top-down predictions have a diffuse connectivity pattern, where every value neuron $\mu_j$ in the layer above affects each error neuron $\epsilon_i$ through the synaptic weight $W_{i,j}$. A one-to-one connectivity structure is a highly precise and sensitive pattern and it is difficult to see how it could first develop and then be maintained in the brain throughout the course of an organisms life. Additionally, while precise connectivity can exist in theory, there is little evidence neurophysiologically \citep{bastos2012canonical,walsh2020evaluating} for the kind of regular and repeatable one-to-one connectivity patterns that predictive coding would require in the brain. Moreover, if predictive coding were implemented throughout the cortex, this one-to-one connectivity should be highly visible to researchers.

One way to address this problem is to formulate predictive coding as a dendritic computation model \citep{guerguiev2017towards,sacramento2018dendritic}, typically postulating that prediction errors are computed on the apical dendrites of each cell. This eliminates the issue of one-to-one connections between value and error neurons by effectively moving the connection within the neuron itself. While current dendritic theories suffer from one-to-one connectivity requirements elsewhere - typically between the pyramidal cells and the inhibitory interneurons, this is not necessarily intrinsic to all possible dendritic error theories. We take a different approach and try to maintain the distinction between value and error neurons, but instead postulate a diffuse connectivity pattern between them, mediated by a set of connection weights $\psi$. The new equation for the prediction errors becomes:
\begin{align}
     &\epsilon_i^l = \sum_k \psi_{i,k}^l \mu_k^l - f(\sum_j \theta^{l+1}_{i,j} \mu^{l+1}_{j}) \\
     &\implies \epsilon^l = \psi^l \mu^l - f(\theta^{l+1} \mu^{l+1})
\end{align}

While using randomly initialized weights $\psi$ completely destroys learning performance, it is possible to learn these weights in an online unsupervised fashion using another Hebbian learning rule. This time the learning rule required is
\begin{align}
    \frac{d\psi^l}{dt} = \epsilon^l {\mu^l}^T
\end{align}
This rule is completely linear and Hebbian since it is simply a multiplication of the activations at the two endpoints of the connection. We show in Figure 6 that using this rule allows equivalent learning performance to the one-to-one case as the required weight values are rapidly learnt.

\begin{figure}[ht] 
  \begin{subfigure}[b]{0.5\linewidth}
    \centering
    \includegraphics[width=0.75\linewidth]{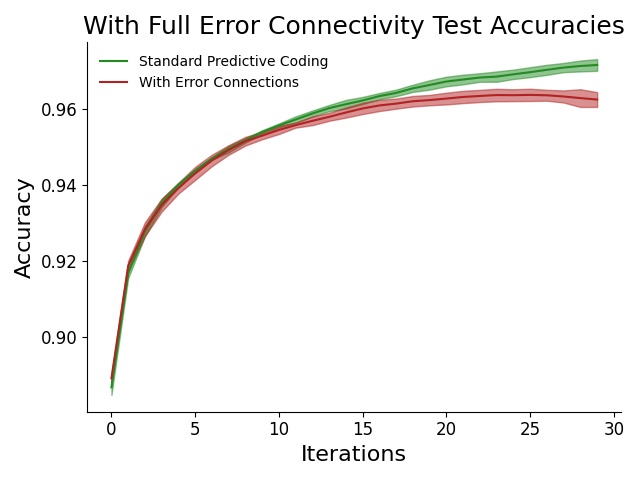} 
    \caption{\small MNIST dataset; tanh activation} 
    \vspace{4ex}
  \end{subfigure}
  \begin{subfigure}[b]{0.5\linewidth}
    \centering
    \includegraphics[width=0.75\linewidth]{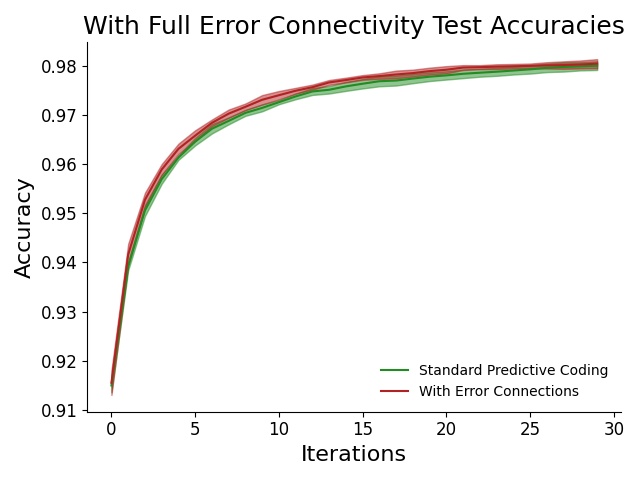} 
    \caption{\small MNIST dataset; relu activation} 
    \vspace{4ex}
  \end{subfigure} 
  \begin{subfigure}[b]{0.5\linewidth}
    \centering
    \includegraphics[width=0.75\linewidth]{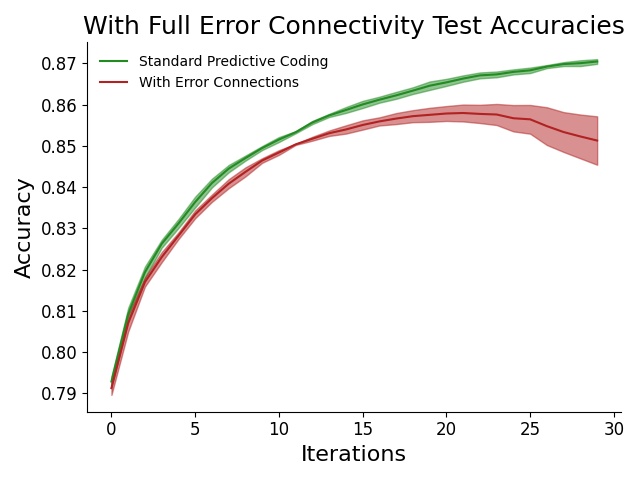} 
    \caption{\small Fashion dataset; tanh activation} 
  \end{subfigure}
  \begin{subfigure}[b]{0.5\linewidth}
    \centering
    \includegraphics[width=0.75\linewidth]{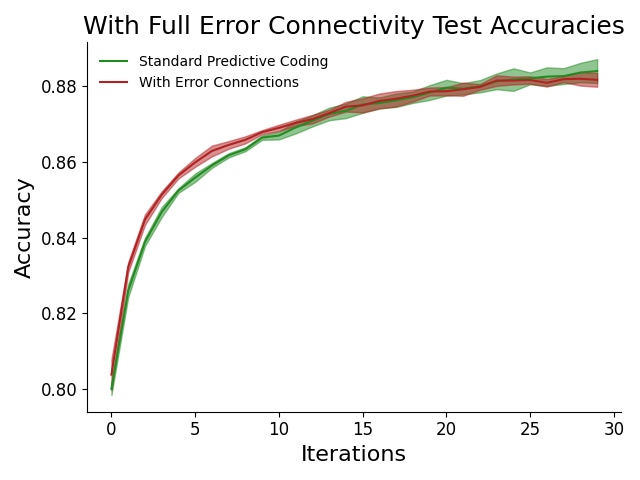} 
    \caption{\small Fashion dataset; relu activation} 
  \end{subfigure} 
  \caption{Test accuracy of predictive coding networks with and without learnable error connections for both relu and tanh activation functions on the MNIST and FashionMNIST datasets. We see that, interestingly, using learnt error weights decreased performance only with the tanh but not the relu nonlinearity.}
\end{figure}

We see that, overall, training performance can be maintained even with learnt error connections, a perhaps surprising result given how key the prediction errors are in driving learning. Interestingly, we see a strong effect of activation function on performance. Performance is indistinguishable from baseline with a relu activation function but asymptotes at a slightly lower value than the baseline with tanh. Investigating the reason for the better performance of the relu nonlinearity would be an interesting task for future work.

\subsection{Combining Relaxations}

It is also possible to combine all of the above relaxations together in parallel, to create a network architecture which is avoids all the major biological plausibility pitfalls of predictive coding. A schematic representation of this combined architecture compared to the standard predictive coding architecture is shown below (Figure 7).

\begin{figure}[H]
    \centering
    \subfloat[\centering Standard PC]{{\includegraphics[width=0.40\textwidth]{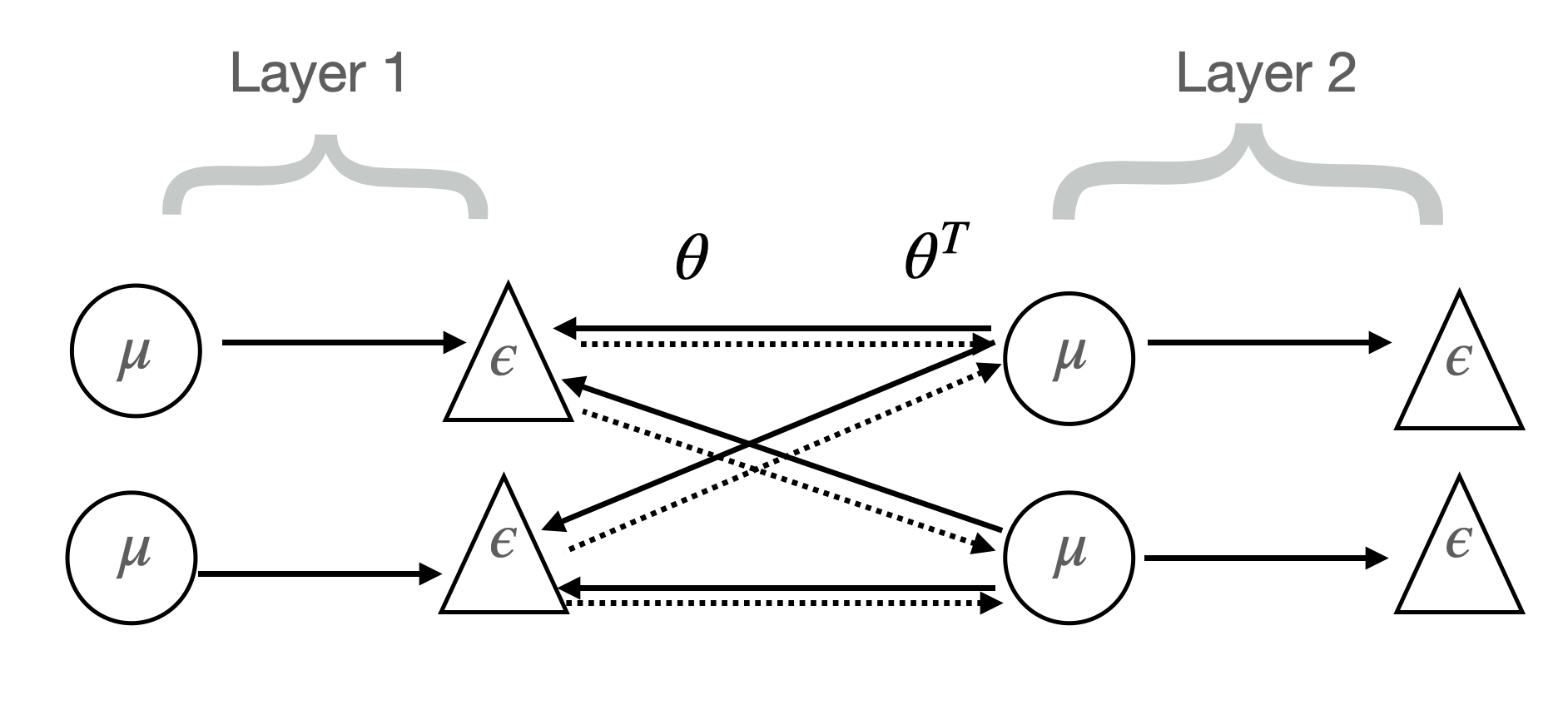} }}%
    \qquad
    \subfloat[\centering Relaxed PC]{{\includegraphics[width=0.40\textwidth]{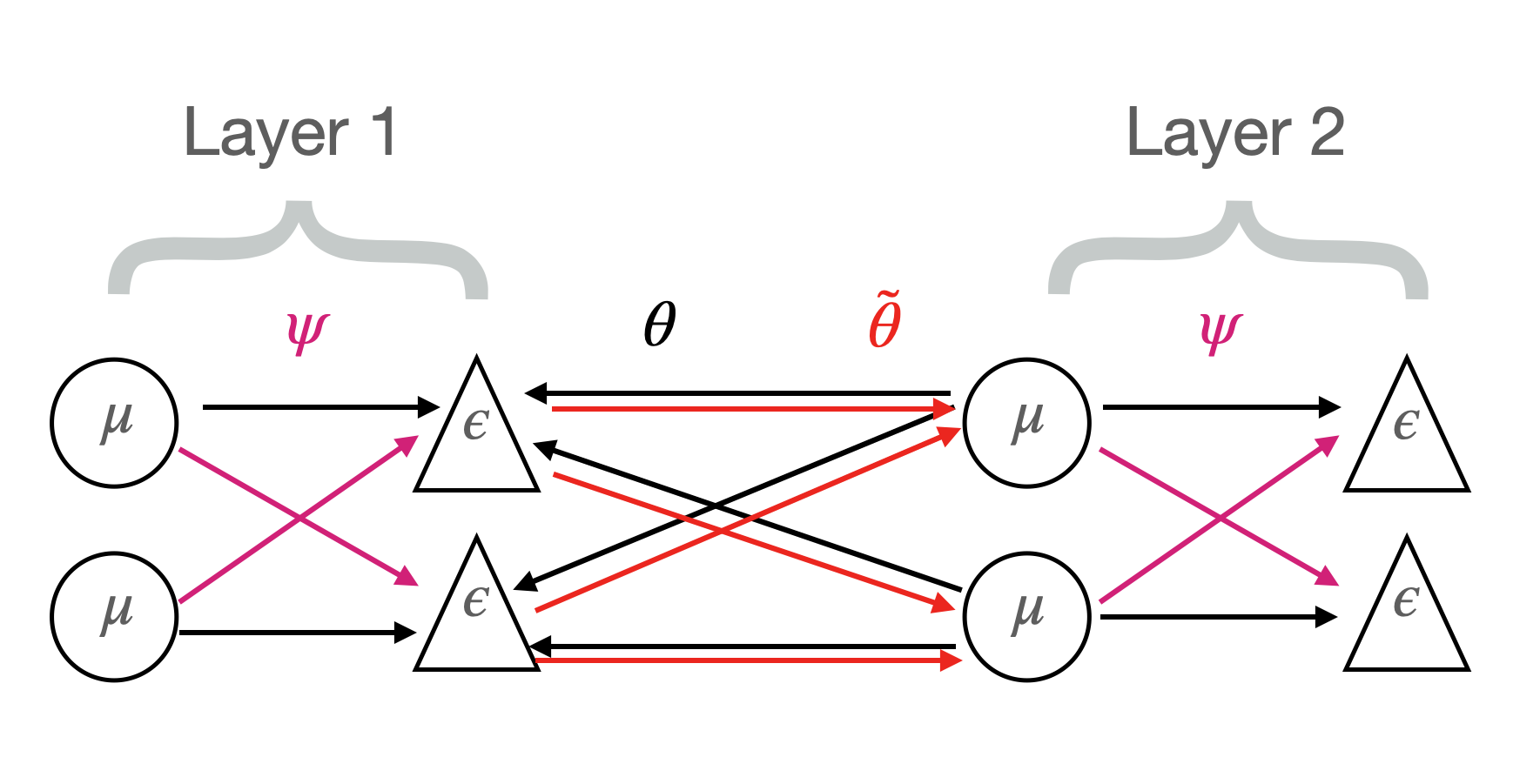} }}%
    \caption{Schematic representations of the architecture across two layers of a.) the standard predictive coding architecture and b.) the fully relaxed architecture. The fully relaxed architecture }%
\end{figure}

\begin{figure}[ht] 
  \begin{subfigure}[b]{0.5\linewidth}
    \centering
    \includegraphics[width=0.75\linewidth]{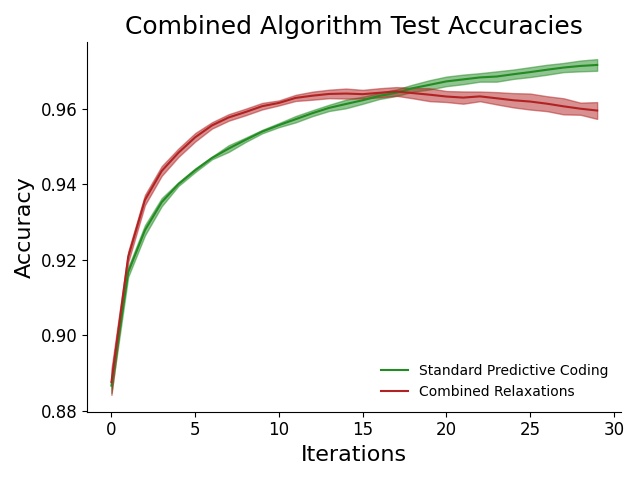} 
    \caption{\small MNIST dataset; tanh activation} 
    \vspace{4ex}
  \end{subfigure}
  \begin{subfigure}[b]{0.5\linewidth}
    \centering
    \includegraphics[width=0.75\linewidth]{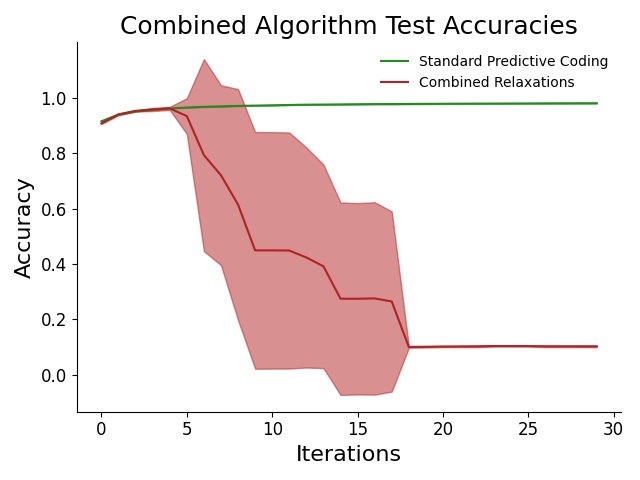} 
    \caption{\small MNIST dataset; relu activation} 
    \vspace{4ex}
  \end{subfigure} 
  \begin{subfigure}[b]{0.5\linewidth}
    \centering
    \includegraphics[width=0.75\linewidth]{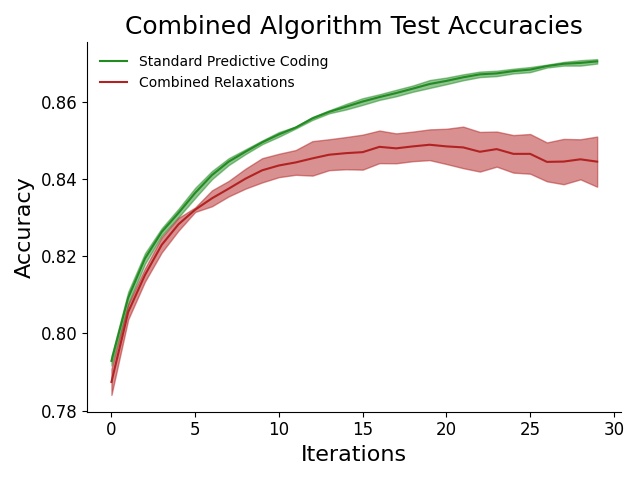} 
    \caption{\small Fashion dataset; tanh activation} 
  \end{subfigure}
  \begin{subfigure}[b]{0.5\linewidth}
    \centering
    \includegraphics[width=0.75\linewidth]{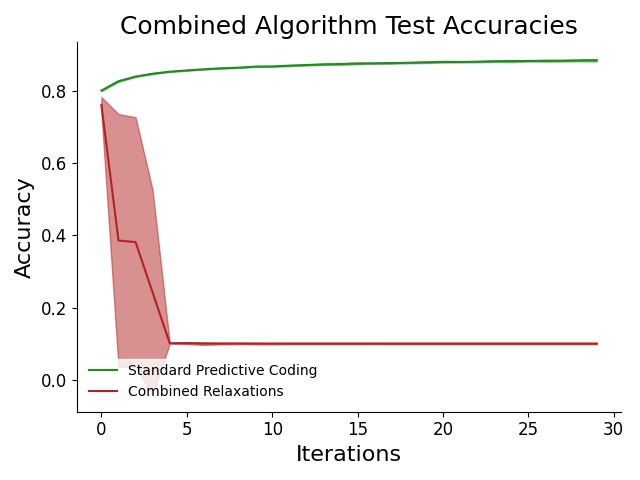} 
    \caption{\small Fashion dataset; relu activation} 
  \end{subfigure} 
  \caption{Test accuracy standard and fully relaxed predictive coding networks, for both relu and tanh activation functions on the MNIST and FashionMNIST datasets. We see that, interestingly, using learnt error weights decreased performance only with the tanh but not the relu nonlinearity.}
\end{figure} 

The relaxed architecture no longer has any 1-1 connectivity patterns, which has been replaced with full connectivity matrices parametrised by the error connection weights $\psi$. Moreover, the forwards and backwards weights are separated into two separate and independent weight matrices $\theta$ and $\tilde{\theta}$ while the standard predictive coding model uses the true weight transpose $\theta^T$, which requires copying the weights. In essence, the fully relaxed architecture simply consists of two bipartite populations of neurons, which only synapse onto the other population. Beyond this there is no special connectivity structure required. Nevertheless, we show that even this very simple architecture, with only Hebbian learning rules, can still be trained to perform well at supervised classification.

We tested the classification ability of the fully relaxed architecture with both hyperbolic tangent and rectified linear activation functions, and on the MNIST and FashionMNIST datasets, and the results are shown in Figure 8. Overall we found that another strong effect of activation function where this time training was unstable and diverged when using rectified linear units but not when using tanh neurons. We hypothesize that this could be due to the rectified linear units not having a saturation point unlike tanh and thus being more prone to exploding gradients. We found, on the other hand, that while performance of the fully relaxed network asymptotically tended to be slightly worse than the standard, it was still very high on both the MNIST and fashion MNIST datasets, thus showing that even highly relaxed and extremely local networks with an extremely generic, essentially fully connected, connectivity pattern can be trained to very high accuracies using this predictive coding algorithm which only requires Hebbian updates. 

\section{Discussion}

We have shown that while there are three outstanding issues of biological plausibility within the predictive coding framework -- the weight transport problem, backward nonlinearities, and the error-connections problem --  it is possible to surmount each problem in a relatively straightforward manner. In the weight transport and error connections case, the solution has been to propose a separate set of weights which are themselves learnable by a Hebbian rule. In the backwards nonlinearities case, it suffices simply to ignore the biologically implausible terms in the rule. Moreover, we have shown that performance, at least under the hyperbolic tangent nonlinearity, is still stable and roughly comparable with the baseline when all the relaxations are combined together, thus resulting in an extremely local,straightforward, and biologically plausible architecture. Overall, we believe these results show that predictive coding offers a surprisingly robust model of learning and inference in the brain and that it can survive often severe perturbations to its basic equations. Moreover, we believe that through this work we have substantially diminished the constraints a predictive coding neuropysiological process theory must satisfy. In so doing, we believe that it is possible that predictive coding may now fit a greater part of neurophysiological data, while opening the way to potentially constructing novel microcircuit designs which implement relaxed forms of predictive coding.
There may also be gains from applying these heuristics to other biologically plausible approximations to backprop. If our heuristics prove successful there, we believe they may hint at general properties of perturbational robustness of these neurally inspired learning algorithms which may be of significant theoretical interest. 

An additional theoretical note is the power of the assumption of variational optimality. Like a Lagrangian in physics, the variational free-energy $\mathcal{F}$ enables potentially complex `laws of motion' to be derived through a simple mathematical apparatus and which can be extended to lead to otherwise-difficult insights. For instance, the learning rules for $\tilde{W}$ and $\psi$ can be derived straightforwardly in the variational framework as simple gradient descents on $\mathcal{F}$ given an augmented generative model containing the $\tilde{W}$ or $\psi$ terms. Regardless of one's theoretical or ontological commitments to variational inference in the brain, the mathematical reformulation of neural activity as encoding solutions to a variational inference problem allows considerable modelling flexibility and mathematical insight through which results can be easily derived which would be much harder to achieve through other means. 

Having removed the symmetric backwards weights and the one-to-one error connectivity scheme, we are left with an essentially bipartite graph. There are connections between the value and error units of the same level, and the error units of one level and the value units of the level above, but crucially there are no direct connections between the value or error units of one layer and those of the layer above. Stepping out of the predictive coding framework, we have effectively shown that a simple bipartite connectivity structure and Hebbian learning rules suffices to learn complex input-output mappings, and may mathematically approximate backpropagation. This is a surprising result given that previously Hebbian learning has not generally been thought to be sufficient to learn complex representations in the brain \citep{baldi2016theory}. This shows that perhaps it is possible for the brain to go further with clever connectivity patterns and Hebbian learning than previously thought.

\section{Acknowledgements}

BM is supported by an EPSRC funded PhD Studentship. AT is funded by a PhD studentship from the Dr. Mortimer and Theresa Sackler Foundation and the School of Engineering and Informatics at the University of Sussex. CLB is supported by BBRSC grant number BB/P022197/1 and by Joint Research with the National Institutes of Natural Sciences (NINS), Japan, program No. 01112005.  AT and AKSare grateful to the Dr. Mortimer and Theresa Sackler Foundation, which supports the Sackler Centrefor Consciousness Science.   AKS is additionally grateful to the Canadian Institute for AdvancedResearch (Azrieli Programme on Brain, Mind, and Consciousness). BM would additionally like to thank Mycah Banks for her invaluable contribution in preparing the figures for this manuscript.

\bibliography{refs.bib}

\section*{Appendix A: Derivation of the Variational Free Energy and the update rules for all parameters}

This appendix is a condensed derivation of the variational free energy used in predictive coding and the assumptions required. For a more didactic treatment of this subject, we recommend \citet{buckley2017free} and \citet{bogacz2017tutorial}.

The goal of variational inference is to approximate an intractable posterior $p(x | o)$ where $o$ are observations of some sort and $x$ are latent states which generated the observations. It is assumed that the generative model $p(o,x)$ is known, while the normalizing constant $p(o)$ is unknown and thus the true posterior $p(x | o) = \frac{p(o,x)}{p(o)}$ cannot easily be computed.

Variational inference tries to approximate the intractable true posterior with a simpler variational distribution $q(x; \mu)$ which is then fit to the true posterior by minimizing the KL divergence between them
\begin{align*}
    \mu^* = \underset{\mu}{argmin} \, \KL \big[ q(x ; \mu) \Vert p(x | o) \big]
\end{align*}
This divergence is intractable to compute directly, since it contains the intractable true posterior term, but we can instead minimize a variational bound on this quantity called the \emph{variational free energy} $\mathcal{F}$ (VFE). 
\begin{align}
   \KL \big[q(x ; \mu) \Vert p(x | o) \big] &= \KL \big[ q(x ; \mu) \Vert \frac{p(o,x)}{p(o)} \big] \\
   &= \KL \big[ q(x ; \mu) \Vert p(o,x) \big] + \ln p(o) \\
&\leq \KL \big[ q(x ; \mu) \Vert p(o,x) \big] \\
   &\leq \mathcal{F}
\end{align}
It is possible to represent the free energy functional as the sum of an energy and an entropy term.
\begin{align}
    \mathcal{F} &= \KL \big[ q(x ; \mu) \Vert p(o,x) \big] \\
    &= - \underbrace{\mathbb{E}_{q(x;\mu)}\big[ \ln p(o,x)  \big]}_{\text{Energy}} - \underbrace{\mathcal{H}\big[ q(x ; \mu) \big]}_{\text{Entropy}}
\end{align}
To proceed further, the form of the generative model $p(o,x)$ and the variational distribution $q(x ; \mu)$ must be defined. To derive predictive coding, we utilize a hierarchical Gaussian generative model for each layer of the network $p(o,x^{1:N}) = \mathcal{N}(o | x^1, \Sigma^1) \prod_{l=1}^N \mathcal{N}(x^l; x^{l+1},\Sigma_i)$. We also utilize a factorised variational distribution $q(x^{1:N}) = \prod_{l=1}^N q(x^l; \mu^l)$. Here, unlike \citep{bogacz2017tutorial} and \citep{buckley2017free}, we do not apply the Laplace approximation to the variational distribution but instead assume it is a dirac-delta $q(x^l ; \mu^l) = \delta(x^l - \mu^l)$ (which represents a point distribution). This simplifies the analysis considerably and leads to effectively the same outcome.

With these distribution defined we can compute the free-energy $\mathcal{F}$. We can ignore the entropy term since in the free-energy, since the entropy of a dirac-delta distribution is negligible (technically negative infinity). This leaves the energy term.
\begin{align}
    \mathcal{F} &= \mathbb{E}_{q(x^{1:N}; \mu^{1:N})}\big[ \ln p(o, x^{1:N}) \big] \\
    &= \int dx^1 q(x^1 ; \mu^1) \ln p(o | x^1) \sum_{l=1}^N \int dx^l q(x^l ; \mu^l) \ln p(x^l | x^{l+1}) \\
    &= \int dx^1 \delta(x^1 - \mu^1) \ln p(o | x^1) \sum_{l=1}^N \int dx^l \delta (x^l - \mu^l)  \ln p(x^l | x^{l+1}) \\
    &= \ln p(o | \mu^1) \sum_{l=1}^N   p(\mu^l | \mu^{l+1}) \big)
\end{align}

Substituting in the definition of the generative model, we thus obtain
\begin{align}
    p(o | \mu^1) \prod_{l=1}^N p(\mu^l | \mu^{l+1}) &= \mathcal{N}(o; \mu^1, \Sigma^1) \prod_{l=1}^N \mathcal{N}(\mu^l; \mu^{l+1},\Sigma^{l+1}) \\
`   &=\ln  \frac{1}{(2\pi)^{\frac{N}{2}} det |\Sigma_1|^{\frac{1}{2}}} e^{(o - \mu^1)^T {\Sigma^1}^{-1} (o-\mu^1)} \sum_{l=1}^N \ln \frac{1}{(2\pi)^{\frac{N}{2}} det |\Sigma^l|^{\frac{1}{2}}} e^{(o - \mu^{l+1})^T {\Sigma^l}^{-1} (\mu^l-\mu^{l-1})} \\
&= \sum_{l=0}^N (\mu^l - \mu^{l-1})^T {\Sigma^l}^{-1} (\mu^l - \mu^{l+1}) + \sum_{l=1}^N \ln \frac{1}{(2\pi)^{\frac{N}{2}} det |{\Sigma^l|}^{\frac{1}{2}}} \\
&= \sum_{l=0}^N \epsilon^l {\Sigma^l}^{-1} \epsilon^l + \sum_{l=0}^N \ln \frac{1}{(2\pi)^{\frac{N}{2}} det {|\Sigma^l|}^{\frac{1}{2}}}
\end{align}
Where $\epsilon^l = \mu^l -f( \theta^{l+1} \mu^{i+1})$ and is the prediction errors, and from step 21 onwards we have defined $o = x_0$ for notational simplicity. Throughout we ignore the variances $\Sigma$ of the generative model, however they can also be learnt. See \citet{bogacz2017tutorial,friston2005theory} for details. From the form of the free energy in Equation 22, it is straightforward to derive the dynamics of the parameters, which are just gradient descents on the free energy. For instance, we derive Equations 2 and 3 as follows 
\begin{align}
    \frac{d\mu^l}{dt} = -\frac{\partial \mathcal{F}}{\partial \mu^l} &= \frac{\partial}{\partial \mu^l}\big[ \sum_{l=0}^N {\epsilon^l}^T {\Sigma^l}^{-1} \epsilon^l \big] \\
    &= \epsilon^l {\Sigma^l}^{-1} \frac{\partial \epsilon^l}{\partial \mu^l} +  \epsilon^{l+1} {\Sigma^{l+1}}^{-1} \frac{\partial \epsilon^{l+1}}{\partial \mu^l} \\
    &= \epsilon^l {\Sigma^l}^{-1} - \epsilon^{l+1} {\Sigma^{l+1}}^{-1} f'(\mu^{l}, \theta^l) {\theta^i}^T
\end{align}
Which is equivalent to Equation 2 if we assume that the variances are the identity matrix, and using the fact that $\frac{\partial \epsilon^l}{\partial \mu^l}  = \frac{\partial (\mu^l - f(\mu^{l-1},\theta^{l-1})}{\partial \mu^l} = 1$ and $\frac{\partial \epsilon_{l+1}}{\partial \mu^l}  = \frac{\partial (\mu^{l+1} - f(\mu^l,\theta^l)}{\partial \mu^l} = f'(\mu^l, \theta^l) {\theta^l}^T$. A similar derivation can be done with the weights $W$ to derive Equation 3.
\begin{align}
    \frac{dW_i}{dt} = -\frac{\partial \mathcal{F}}{\partial \theta^l} &= \frac{\partial}{\partial \theta^l}\big[ \sum_{l=0}^N {\epsilon^l}^T {\Sigma^l}^{-1} \epsilon^l \big] \\
    &= \epsilon^{l+1} {\Sigma_{l+1}}^{-1} \frac{\partial \epsilon^{l+1}}{\partial \theta^l} \\ 
    &= - \epsilon^l{l+1} {\Sigma^l}^{-1} f'(\mu^{l}, \theta^l) {\mu^l}^T
\end{align}

To derive the update rule for the error connection matrix $\psi$ (Equation 8), we redefine prediction errors to be equal to $\epsilon^l = \psi^l \mu^l -  f(\mu_{l-1}, W_{l-1})$. We can then compute the gradient of the free-energy with respect to this new parameter $\psi$.
\begin{align}
    \frac{d\psi}{dt} = -\frac{\partial \mathcal{F}}{\partial \psi} &= \frac{\partial}{\partial \psi}
 \big[ \epsilon^l {\Sigma^l}^{-1} \epsilon^l \big] \\
 &= \epsilon^l  {\Sigma^l}^{-1} \frac{\partial \epsilon^l}{\partial \psi} \\
 &=\epsilon^l  {\Sigma^l}{-1} \frac{\partial (\psi^l \mu^l - f(\mu^{l+1},\theta^l))}{\partial \psi} \\
 &= \epsilon^l  {\Sigma^l}^{-1} {\mu^l}^T
 \end{align}
 
 which is identical to Equation 8.
\end{document}